%% file: main.tex
\documentclass[manuscript, nonacm]{acmart}
\AtBeginDocument{%
  }

\setcopyright{acmlicensed}
\copyrightyear{2018}
\acmYear{2018}
\acmDOI{XXXXXXX.XXXXXXX}
\acmConference[Conference acronym 'XX]{Make sure to enter the correct
  conference title from your rights confirmation email}{June 03--05,
  2018}{Woodstock, NY}
\acmISBN{978-1-4503-XXXX-X/2018/06}

\usepackage{longtable}
\usepackage{array}
\usepackage{upquote}
\usepackage[utf8]{inputenc}
\usepackage[dvipsnames]{xcolor}
\usepackage{booktabs, graphicx, array}
\usepackage{tabularx}
\usepackage{soul}

 	\definecolor{amber}{rgb}{1.0, 0.49, 0.0}

\definecolor{grainColor}{HTML}{FF0000}
\definecolor{gapColor}{HTML}{B000BA}
\definecolor{horizonColor}{HTML}{284878}
\definecolor{explicitColor}{HTML}{406828}
\definecolor{provenanceColor}{HTML}{E09848}
\definecolor{commitmentColor}{HTML}{60A098}

\colorlet{grainHL}{grainColor!15!white}
\colorlet{gapHL}{gapColor!15!white}
\colorlet{horizonHL}{horizonColor!15!white}
\colorlet{explicitHL}{explicitColor!15!white}
\colorlet{provenanceHL}{provenanceColor!15!white}
\colorlet{commitmentHL}{commitmentColor!15!white}

\newcommand{\bemph}[1]{\emph{\textbf{#1}}}

\newcommand{\dimvalue}[1]{\par\noindent\textcolor{black}{\textbf{#1}}:}

\begin{document}

\title{How Systems Treat Intent: A Faceted Taxonomy of Intent for Human-AI Co-Creation}

\author{Ashwin Ram}
\authornote{Equal contribution}
\affiliation{%
  \institution{HCI Lab, Saarland University, Saarland Informatics Campus}
  \city{Saarbrücken}
  \country{Germany}
}
\email{ram@cs.uni-saarland.de}

\author{Radha Kumaran}
\authornotemark[1]
\affiliation{%
  \institution{HCI Lab, Saarland University, Saarland Informatics Campus}
  \city{Saarbrücken}
  \country{Germany}
}
\email{kumaran@cs.uni-saarland.de}

\author{Alexander Koller}
\affiliation{%
  \institution{Department of Language Science and Technology, Saarland University}
  \city{Saarbrücken}
  \country{Germany}
}
\email{koller@coli.uni-saarland.de}

\author{Kevin Baum}
\affiliation{%
  \institution{Center for European Research in Trusted AI (CERTAIN), German Research Center for Artificial Intelligence (DFKI)}
  \city{Saarbrücken}
  \country{Germany}
}
\email{kevin.baum@dfki.de}

\author{Jürgen Steimle}
\affiliation{%
  \institution{HCI Lab, Saarland University, Saarland Informatics Campus}
  \city{Saarbrücken}
  \country{Germany}
}
\email{steimle@cs.uni-saarland.de}

\renewcommand{\shortauthors}{Ram \& Kumaran et al.}

\begin{abstract}

Human-AI interaction depends on systems acting on user intent rather than explicit procedures, even when that intent is incomplete, evolving, or partially communicable. Literature organizing systems by interaction technique, workflow stage, or division of labor, offer limited insight into which facet of intent an interaction mechanism addresses. We introduce a six-facet taxonomy of intent, grounded in philosophical accounts of mind, action, and meaning, and apply it to systems presented in 198 papers from eleven SIGCHI venues. Our analysis reveals that different forms of incomplete intent are associated with distinct interaction and authorship strategies, showing that mechanisms that appear similar often address fundamentally different intent-related problems. This exposes blind spots, as intent is rarely represented as an interaction object, authorship provenance is rarely tracked, and tentative preferences are rarely distinguished from firm commitments. The taxonomy offers a lens for analyzing and designing AI systems around the facets of intent they address.

\end{abstract}

\ccsdesc[500]{Do Not Use This Code~Generate the Correct Terms for Your Paper}
\ccsdesc[300]{Do Not Use This Code~Generate the Correct Terms for Your Paper}
\ccsdesc{Do Not Use This Code~Generate the Correct Terms for Your Paper}
\ccsdesc[100]{Do Not Use This Code~Generate the Correct Terms for Your Paper}

\keywords{Human-AI interaction, user intent, interaction mechanisms, taxonomy, generative AI, mixed-initiative interaction, systematic literature review.}

\begin{teaserfigure}
    \centering
    \includegraphics[width=\textwidth,height=0.35\textheight,keepaspectratio]{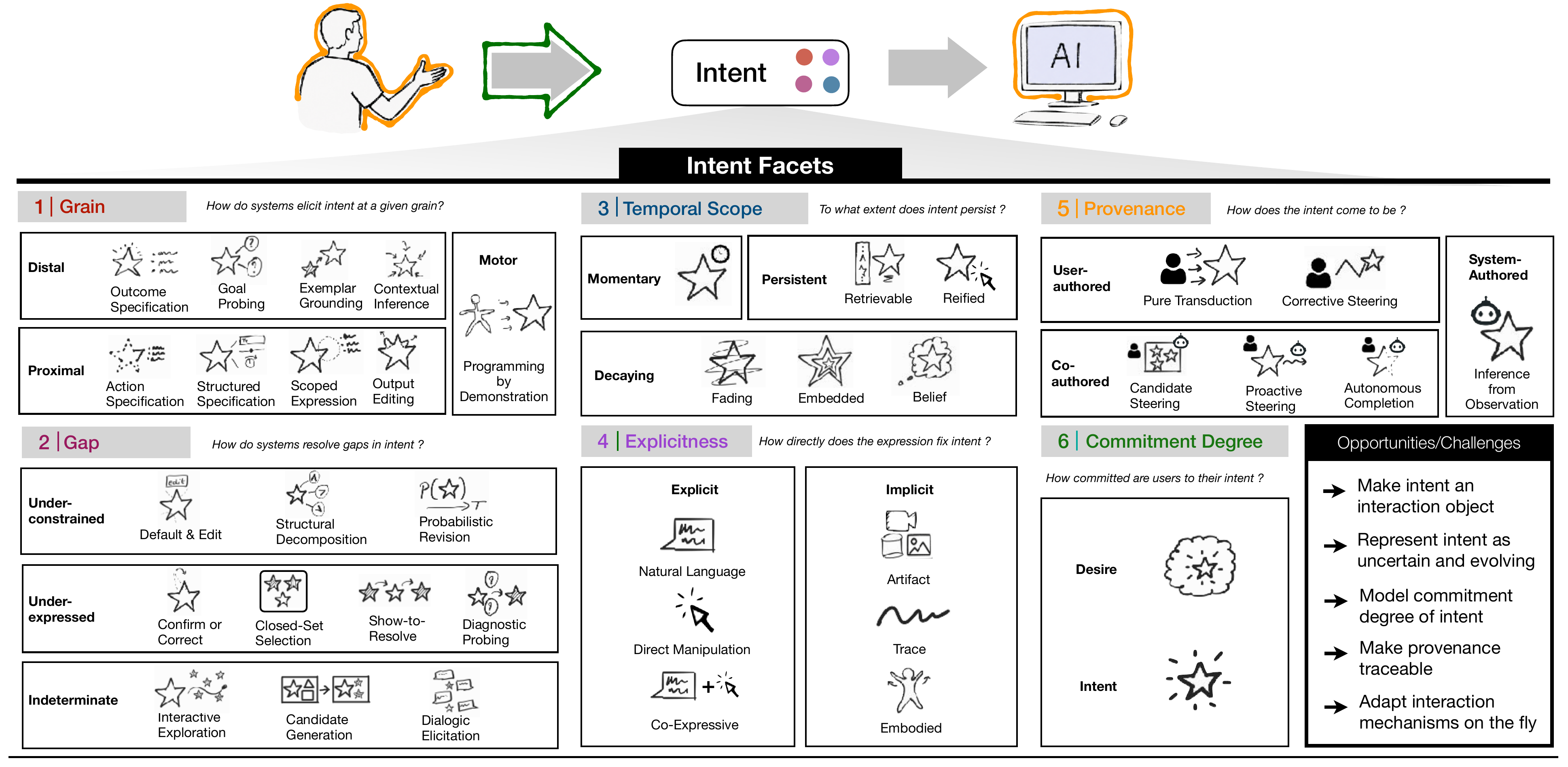}
    \caption{Taxonomy of intent in human-AI interaction. It organizes interactive AI systems according to six facets of intent, the values each facet can take, and the interaction mechanisms through which systems elicit, represent, and act on those values. By relating interaction mechanisms to facets of intent, the taxonomy enables systematic comparison not only of what systems do, but of the intent-related problems they are designed to resolve and the assumptions they make about users. }
    \label{fig:taxonomy}
\end{teaserfigure}


\maketitle

\input{sections/1.intro}
\input{sections/2.related_work}

\input{sections/3.theory-dimensions}

\input{sections/4.SLR}

\input{sections/5.analysis}
\input{sections/6.discussion}

\input{sections/7.future}
\input{sections/8.conclusion}

\bibliographystyle{ACM-Reference-Format}
\bibliography{sample-base}



\end{document}

%% file: sections/1.intro.tex
\section{Introduction}

Understanding and operationalizing user intent is a longstanding challenge in human-computer interaction. Norman's notion of the gulf of execution describes the distance between a user's intentions and the actions a system makes available to express them \cite{norman2013design}. Recent AI systems appear to change how this gulf is crossed. Rather than requiring users to specify procedures step by step, systems increasingly ask users to specify desired outcomes while determining the means of achieving them autonomously. As systems assume greater responsibility for deciding \textit{how} to act, interaction depends increasingly on how effectively they can act on \textit{what} users want. 

This dependence is not new. Interactive systems have long inferred user intent through approaches such as plan recognition, programming by demonstration, and mixed-initiative interaction~\cite{allen1980analyzing, lieberman2001your, horvitz1999mixed}. In these systems, however, the space of intentions that could be recognized was typically bounded by predefined plans, actions, or representations. Contemporary human-AI systems substantially relax this constraint. Users can express goals, preferences, and constraints that need not correspond to any predefined action model, leaving systems to interpret requests whose meaning is often incomplete, evolving, or only partially communicable. Consequently, the challenge is not simply one of more accurate intent recognition. In many cases there is no complete intent representation against which recognition can succeed. Uncertainty about what the user wants becomes a residual property of interaction rather than a transient recognition error, and systems must decide how to proceed despite it.

Different systems address this uncertainty through different interaction mechanisms. Faced with the same underspecified request, one system may generate a proposed interpretation for the user to revise, while another may present several alternatives from which to choose. Although both address incomplete intent, they make different assumptions about what the user knows and how uncertainty should be resolved. They also fail differently when those assumptions do not hold. Interaction mechanisms therefore do more than facilitate communication. They embody particular assumptions about the relationship between what a user has expressed and what the system still needs to infer.

The central premise of this paper is that \textbf{understanding human-AI interaction requires examining not only how users communicate with systems, but also the properties of intent that interaction mechanisms are designed to address}. Mechanisms that appear similar at the interface may serve fundamentally different purposes because they respond to different conditions of intent. For example, systems that ask follow-up questions may help users state an existing preference, discover a preference that is not yet fully formed, or provide information that has not yet been communicated. Although the interaction is similar in each case, the underlying intent-related problem is different. Conversely, systems may employ very different interaction techniques to address the same underlying facet of intent. We therefore argue that interactive mechanisms differ not only in how they communicate with users, but also in the properties of intent they presume, elicit, and act upon. Understanding those properties is essential for explaining why particular mechanisms are appropriate, when they succeed, and how they fail.

Existing reviews have organized interactive AI systems according to interaction modalities, workflow stages, support functions, and divisions of labor between users and AI systems. These perspectives provide valuable accounts of what systems enable users to do. However, they offer less support for a complementary design question: \textit{what facets of intent are particular mechanisms designed to address?} In doing so, we propose intent facets as a complementary organizing principle for human-AI interaction, enabling mechanisms to be compared according to the conditions they address rather than solely the interactions they perform.

To support this analysis, we conducted a systematic review of 198 interactive human-AI systems published across eleven SIGCHI venues, identifying distinct interface mechanisms through which systems elicit, represent, and act on user intent. To compare these mechanisms, we draw on philosophical accounts of mind, action, and meaning to characterize intent along six facets: Grain, Gap, Temporal Scope, Explicitness, Provenance, and Commitment Degree. These facets provide an analytical vocabulary for describing key properties of intent and relating them to the mechanisms through which systems establish what users want. By connecting intent facets to interaction mechanisms, we construct a taxonomy  (see Figure ~\ref{fig:taxonomy}) that reveals how different mechanisms address different conditions of intent and makes it possible to systematically compare systems that would otherwise appear unrelated.

Our analysis reveals recurring patterns that are difficult to see from an interaction-centered perspective alone. Mechanisms that appear similar at the interface often address fundamentally different intent-related problems: helping users discover preferences that are not yet fully formed, specify details that remain open, or clarify intentions that have already been decided but not successfully communicated. These differences also shape how authorship is distributed between users and AI systems, suggesting that patterns of user and system control emerge from the kind of intent being resolved rather than from the interaction technique itself. More broadly, while systems frequently participate in shaping user intent, intent itself is rarely represented as an explicit object that users can inspect, revise, and reuse. Likewise, co-authored intent is seldom traced back to the respective contributions of users and systems, and current systems almost never distinguish tentative preferences from committed intentions. Together, these patterns expose important blind spots in how human-AI systems represent and act on intent, and suggest opportunities for future interaction design, which we discuss in the paper.

This work makes three contributions:

\begin{enumerate}
    \item \textbf{An inventory of interface mechanisms} through which interactive AI systems elicit, represent, and act on user intent, derived from a systematic review of 198 papers.
    \item \textbf{A taxonomy relating intent facets to interaction mechanisms}, enabling systematic comparison of how different mechanisms address different forms of user intent and the assumptions they make about users.
    \item \textbf{An analysis of patterns, blind spots, and opportunities within this taxonomy}, revealing how different forms of incomplete intent are associated with distinct interaction and authorship strategies, and identifying underexplored opportunities for representing, tracking, and managing intent in human-AI interaction.
\end{enumerate}

Together, these contributions provide a basis for reasoning about the fit between interaction mechanisms and user intent, enabling researchers and designers to evaluate not only whether a mechanism works, but whether it is appropriate for the kind of intent a system is expected to support. The paper reframes interactive AI systems as responses to different conditions of intent. Rather than organizing systems by interaction workflows or stages of intent communication, it becomes possible to compare them by the inherent facets of intent they are designed to address, the assumptions they make about users, and the intent-related problems they are able to resolve.

%% file: sections/2.related_work.tex
\section{Related Work}
\label{sec:relatedwork}

\subsection{Intent in Interactive Systems}
\label{sec:intent-history}

Norman's notion of the gulf of execution names the distance between a user's intentions and the actions a system makes available to express them \cite{norman2013design}, and closing this gulf has motivated much of the subsequent history of interactive systems.
Suchman further argues that plans do not by themselves determine situated action: a plan functions as a resource agents draw on and depart from as circumstances unfold, rather than as a specification that action merely executes~\cite{suchman1987plans}. Interactive systems have nonetheless long acted on inferred user intent. Plan-based dialogue systems recognized a speaker's plan from a single utterance and used it to generate cooperative responses that went beyond what was literally requested~\cite{allen1980analyzing}; programming-by-example systems let users demonstrate a task so that a system could generalize a procedure from the demonstration rather than have it specified in code~\cite{lieberman2001your}; and mixed-initiative interfaces interleaved user and system contributions under an explicit model of who should act and when~\cite{horvitz1999mixed}. In each of these approaches, however, the space of intentions a system could recognize was bounded: a plan library the system could match against, a demonstrated trace it could generalize from, or a predefined action model it could reason over. Underspecification was therefore a transient recognition problem, resolved by matching an observed action against an antecedently known structure, rather than a standing condition a design had to take a stance on. Contemporary human-AI systems relax this bound substantially. Users can express outcomes, preferences, and constraints that need not correspond to any predefined plan or action model, and a system frequently has no enumerable structure to recognize against in the first place.

\subsection{Surveys of Interactive AI Systems}
\label{sec:reviews}

A growing body of review work has organized this space by what users do with generative AI systems. At the broadest level, Shi et al.\ survey 291 papers to characterize the purposes for which people use generative AI, the feedback systems return to users, and application domains and evaluation strategies~\cite{shi2024genaitaxonomy}, while Gao et al.\ identify phases of human-LLM interaction and a corresponding taxonomy of interaction modes~\cite{gao2024taxonomy}. Other reviews target specific domains including organizing writing assistance by task, user, technology, interaction, and ecosystem~\cite{lee2024writing} and co-creative workflows by the division of creative labor between user and system~\cite{rezwana2023cofi}. Hu et al.\ reviewed generative AI for art and creative making and identify four paradigms -- medium, tool, partner, and mediator -- characterizing the relational stance a system takes toward the user~\cite{hu2025designing}, and Park et al.\ reviewed image generation tools by interaction method, creative phase, and tool functionality~\cite{park2026interactionmethods}. Across this body of work, the organizing unit is the interaction: the channel through which input arrives, the phase of the workflow, or the division of labor between user and system, rather than the intent that interaction addresses. These reviews give a valuable account of what interactive AI systems currently offer, but offer limited guidance on which offering fits which kind of intent.

\subsection{Intent as the Unit for Analysis}
\label{sec:intent-analysis}

A smaller body of work takes users' intent itself, rather than the interaction, as the unit of analysis. Subramonyam et al.\ extend Norman's stages of action to prompt-based interaction and characterize a gulf of envisioning, comprising gaps in knowing what the task should be, how to instruct the model, and what to expect of the output~\cite{subramonyam2024gulf}. Their account is cognitive: it explains the difficulty a user faces in forming an intent, and centers on characterizing that difficulty rather than on how a system takes up and acts on the intent it receives.

Kim et al.\ conduct a systematic review of 46 papers and identify four aspects of intent communication support, articulation, exploration, management, and synchronization, which they instantiate jointly in a research probe~\cite{kim2026intentflow}. Their work shares much of our motivation, but decomposes a different object: the \emph{process} of communicating an intent, into the stages a user moves through as it is articulated, explored, and kept aligned with the system's understanding. We decompose \emph{intent itself} into properties that characterize it at any point in that process. The two cut across one another, since mechanisms grouped together as supporting articulation may address quite different properties of intent. Where Kim et al.\ ask what support a system provides at each stage of intent communication, we ask what property of intent a mechanism presupposes, which is what determines whether a mechanism suits the intent a system must handle, and how it fails when it does not.

The vocabulary we use to specify these properties, and the corpus against which we ground it, are introduced next.

%% file: sections/3.theory-dimensions.tex
\section{Methodology}

\subsection{A Faceted Account of Intent}
\label{sec:facets}

Intention has a much longer history in philosophy than the recent prominence of intent in AI interaction might suggest. We draw on three bodies of work, the philosophy of mind, the philosophy of action, and the theory of meaning as an analytical lens for structuring our coding scheme. We set facets in bold and their values in italics within parentheses.

On a standard analysis in the philosophy of mind, an intentional state consists of a content held in a specific attitude or psychological mode~\cite{searle1983intentionality}. The content fixes the state's conditions of satisfaction: what would have to be the case for a belief to be true, a desire fulfilled, or an intention carried out. The same content can be held in different modes, as when one believes that it will rain or hopes that it will. Desire and intention are the modes relevant here: unlike belief, both have a world-to-mind direction of fit; the world is to match them, rather than they the world. We use \emph{intent} for the content side: the condition that a user's desire or intention places on what a system is to bring about.\footnote{Strictly, a desire and an intention do not share their full content: on Searle's account, an intention represents the agent's own action and is causally self-referential~\cite{searle1983intentionality}. What they have in common is the condition they place on the outcome, and this condition is what we call intent.} An intent is therefore not itself an intention, and it can stem from a mere desire as well as from a settled intention.

Intentional states also have a bearer (or subject). When agents act on their own intentions, bearer and executor coincide, and the bearer can be left implicit. In human-AI interaction, however, it cannot: a system may act on the user's intent, on goals it pursues on its own account, or on an intent attributed to the pair. Our corpus is scoped to the first case where the user bears the intent, and the system acts on it. 


Because bearer and executor come apart, every system design must take a stance on three questions that give rise to the six facets described below in more detail. The first, what the system takes the user to want, is structured by the analysis above: \textbf{Grain} concerns the level at which the intent is pitched, \textbf{Gap} whether it settles the details the system must decide, and \textbf{Commitment} the attitude from which it is presumed to stem. The other two concern the separation itself: how the intent enters the interaction, addressed by \textbf{Explicitness} (how directly it is expressed) and \textbf{Provenance} (who shapes it), and how the system carries it forward, addressed by \textbf{Temporal Scope}.
  

\paragraph{Provenance}
The intent a system acts on can be shaped by the user, by the system, or by both. Since we code the intent as the system construes it (Section~\ref{sec:analysis}), provenance concerns who determines that construal, not how the user's own attitudes arose. The user may determine it alone, however much the system generates in carrying it out (\textit{User-authored}); the system may determine it alone (\textit{System-authored}); or both may contribute, as when the user selects among or builds on what the system proposes (\textit{Co-authored}). We term this facet \textbf{Provenance}. It is related to, but distinct from, initiative in mixed-initiative interaction~\cite{horvitz1999mixed}: a system can take the initiative without shaping the intent, as when it asks a clarifying question, and shape the intent without taking the initiative, as when it silently fills in a default.

\paragraph{Grain}

Intentions operate at different levels. Pacherie~\cite{pacherie2008phenomenology} distinguishes distal intentions, formed in advance of acting, proximal intentions, which anchor an action in the current situation, and motor intentions, which guide its sensorimotor execution. The levels differ in how far their content determines the way the action is to be performed. An intent may specify an end state and leave to the system how to reach it (\textit{Distal}); specify a concrete operation or parameter setting, so that the way to the outcome is largely settled (\textit{Proximal}); or specify an action at the level of its physical performance, the movements, clicks, or keystrokes the system is to reproduce or generalize from (\textit{Motor}). We term this facet \textbf{Grain}. A motor-grain intent is not a motor intention in Pacherie's sense: it does not guide the user's own movements but specifies, at the level of movement, what the system is to do.

\paragraph{Gap}
However much a user conveys, a system usually has to decide details the user has not specified. What this requires depends on what the design presupposes about the intent in these respects. Its conditions of satisfaction may not discriminate among the ways a detail could be settled, so that any reasonable choice satisfies it (\textit{Underconstrained}). They may discriminate, because the user has a specific answer in mind that has not been conveyed (\textit{Underexpressed}). Or they may not yet discriminate although the user is not indifferent: faced with alternatives, the user would accept some and reject others, forming the missing condition in the process, so that the interaction is where the detail gets settled (\textit{Indeterminate}). The first and the third case reflect what Bratman calls the partiality of intention: an agent can commit to an end without settling every detail of how it will be achieved, filling in the plan as action proceeds~\cite{bratman1987intention}. What separates them is whether the detail is one the user needs to settle at all. The second case is a gap not in the intent but in its expression. We term this facet \textbf{Gap}; designs that presuppose that nothing relevant remains open are coded as having no gap.


\paragraph{Temporal Scope}
Intentions are not momentary. On Bratman's account, a settled intention persists and structures subsequent conduct until it is executed or reconsidered~\cite{bratman1987intention}, and personal policies are intentions meant to govern recurring situations indefinitely~\cite{bratman1989personalpolicies}. Cohen and Levesque~\cite{cohen1990intention} formalize this persistence, characterizing an intention as a goal an agent maintains until it believes the goal achieved, unachievable, or no longer relevant. A user's desire or intention may thus outlast the turn in which its intent is expressed. Whether the system keeps acting on the intent depends on how it carries the intent forward, which we term \textbf{Temporal Scope}. The system may keep no representation of it beyond the current turn (\textit{Momentary}); keep it available while it progressively loses influence over subsequent behavior (\textit{Decaying}); or carry it forward without loss of influence until it is revised or revoked (\textit{Persistent}).



\paragraph{Explicitness}
Theories of meaning distinguish what an utterance says from what a hearer must infer beyond it~\cite{grice1975logic}. Determining what is said can already require context, for instance to settle what ``it'' refers to, but not working out what the speaker wants beyond that. We extend this distinction by analogy to expression in any modality. In some cases, the intent's conditions of satisfaction can be read off the expression itself, as with a stated request, a slider setting, or a pointed-at object (\textit{Explicit}). In others, substantial parts of them must be inferred beyond what the user provides, as with examples or partial sketches, and at the limit from behavior not addressed to the system at all, which the system reads as a sign of what the user wants rather than as communication (\textit{Implicit}). We term this facet \textbf{Explicitness}. Implicit expression is common because much of what users want is tacit: as Polanyi~\cite{polanyi1966tacit} put it, ``we can know more than we can tell,'' and a user may recognize a satisfactory outcome without being able to state what makes it so. Tacitness explains why intent is so often conveyed implicitly; the facet codes the expression itself.

\paragraph{Commitment Degree}
The intent a system acts on stems from some attitude of the user. Bratman~\cite{bratman1987intention} argues that intention is not simply a strong desire but a distinct attitude. Desires are merely potential influencers of conduct, to be weighed against one another; intentions are conduct-controlling: they settle what the agent will do, resist reconsideration, and constrain further planning. The same distinction underlies the BDI tradition in agent architectures~\cite{cohen1990intention,rao1991modeling}. It motivates the facet \textbf{Commitment}: whether the design treats the intent as stemming from a desire, which the system may weigh against other considerations, partially satisfy, or set aside (\textit{Desire}), or from an intention, whose intent the system is expected to realize (\textit{Intention}). Designs that treat all input alike are coded as not differentiating.

Table~\ref{tab:facet-derivation} summarizes the facets and their values. We do not claim that they exhaust what the philosophy of mind and action has to say about intentional states; they isolate the respects in which the systems in our corpus observably differ in how they treat intent. We now present the corpus and coding procedure.

\begin{table}[t]
\small
\renewcommand{\arraystretch}{1.3}
\begin{tabularx}{\textwidth}{@{}
  >{\raggedright\arraybackslash}p{0.19\textwidth}
  l
  >{\raggedright\arraybackslash}X
  >{\raggedright\arraybackslash}p{0.26\textwidth}@{}}
\toprule
\textbf{Question} & \textbf{Facet} & \textbf{What it codes} & \textbf{Values} \\
\midrule
What does the user want?
  & Grain & level at which the intent is pitched & Distal, Proximal, Motor \\
  & Gap & what the design presupposes about unspecified details & Underconstrained, Underexpressed, Indeterminate (or no gap) \\
  & Commitment & attitude from which the intent is presumed to stem & Desire, Intention (or not differentiated) \\
\midrule
How does the intent enter the interaction?
  & Explicitness & how directly the expression fixes the intent's conditions of satisfaction & Explicit, Implicit \\
  & Provenance & who determines the intent the system acts on & User-authored, System-authored, Co-authored \\
\midrule
How is it carried forward?
  & Temporal Scope & how the system represents the intent over subsequent interaction & Momentary, Decaying, Persistent \\
\bottomrule
\end{tabularx}
\caption{The six facets and their values, grouped by the question each addresses.}
\Description{The table groups the six facets by the question they address. What the user wants: Grain, Gap, Commitment. How the intent enters the interaction: Explicitness, Provenance. How it is carried forward: Temporal Scope. For each facet, it states what the facet codes and lists its values.}
\label{tab:facet-derivation}
\end{table}

%% file: sections/4.SLR.tex
\subsection{Systematic Literature Review Method}
\label{sec:method}

We conducted a systematic literature review to construct a representative corpus of interactive AI systems, coded against the facets developed in Section~\ref{sec:facets}. Our goal was to document intent communication with interactive systems using a representative sample of research, not an exhaustive scoping review, so our search terms, venue selection, and eligibility criteria reflect this choice.

\subsubsection{Dataset and Exclusion Criteria}
\label{sec:dataset}


To collect a representative set of interactive AI systems, we conducted a systematic search in the ACM Digital Library, restricted to SIGCHI flagship and affiliated venues, namely CHI, UIST, IUI, DIS, CSCW, HRI, UMAP, Creativity \& Cognition, CUI, TOCHI, and PACMHCI. Our search was restricted to papers published between January 2016 and August 2026 and tagged with one or more of the following ACM CCS 2012 concepts: Interactive systems and tools, Interaction techniques, Interaction paradigms, Natural language interfaces, Collaborative interaction, and Collaborative content creation. After removing records that were not conference or journal papers, this gave us a total of 1{,}352 unique records. Two authors independently screened the titles and abstracts of these records against our exclusion criteria (Table~\ref{tab:eligibility-criteria}). Both authors excluded the same 977 records, which were removed without discussion. The full texts of the remaining 375 papers were then reviewed in a second round, and a further 182 papers were then excluded, with disagreements resolved through discussion. To complement this keyword search, we also identified an additional 5 relevant papers by leveraging the authors' expertise in intent communication and human-AI interaction, papers that satisfied all inclusion criteria but were not tagged with the expected CCS concepts. By merging these papers, we finally selected a corpus of 198 papers for our literature review.

Our exclusion criteria Table~\ref{tab:eligibility-criteria} are intended to focus the corpus on interactive AI systems in which user intent is meaningfully communicated and can shape an open-ended outcome. EC1–EC3 exclude work without a concrete user-facing system, without meaningful variation in user goals or preferences, or where AI does not meaningfully engage with what the user is trying to accomplish. EC4 further scopes the corpus to open-ended generative and co-creative tasks, such as writing, coding, and design. This focus is meaningful for our analysis because these tasks provide a setting in which users may communicate under-specified, evolving, or multi-faceted intentions, and where AI must interpret and act on these intentions to form or transform an outcome.

Our systematic compilation provides broad coverage of research on intent communication in interactive AI systems, but it cannot be considered a complete or exhaustive representation of the space. The boundaries of the corpus are necessarily shaped by our inclusion and exclusion criteria, and some papers may fall near these boundaries or reflect design choices that are difficult to categorize cleanly. Our goal, however, is not to exhaustively enumerate all interactive AI systems, but to develop a taxonomy grounded in the systems represented in the literature. We therefore view the resulting corpus as a representative basis for identifying recurring facets and mechanisms of intent communication within our defined scope. To support further refinement and extension of the taxonomy, we make our coded corpus openly available.

\begin{table*}[t]
  \caption{Exclusion criteria (EC) applied during screening.}
  \label{tab:eligibility-criteria}
  \begin{tabular}{@{}lp{0.85\textwidth}@{}}
    \toprule
    \textbf{\#} & \textbf{Criterion} \\
    \midrule
    EC1 & Excludes surveys, reviews, workshops, position papers, conceptual frameworks, and technical model papers that introduce no concrete user-facing system, prototype, or interaction technique. \\
    EC2 & Excludes systems whose interaction is limited to fixed commands or predefined inputs, allowing no meaningful variation in users' goals, or desired outcomes. \\
    EC3 & Excludes systems where GenAI mainly supplies feedback, recommendations, analyses, or visualizations without meaningfully engaging with what user's intent. \\
    EC4 & Restricts scope to open-ended generative or co-creative tasks (writing, coding, design) where GenAI forms, transforms, or produces a task outcome from communicated or inferred intent; excludes AI supporting understanding, analysis, monitoring, configuration, or operational control without contributing to an open-ended outcome. \\
    \bottomrule
  \end{tabular}
\end{table*}

\subsection{Analysis}
\label{sec:analysis}

The dataset was analyzed through a multi-step, hybrid deductive and inductive process. Facets and their values grounded in the theoretical framework of Section \ref{sec:facets} were treated as an a priori coding scheme and applied deductively to every paper in the corpus. For the interface mechanisms addressing each facet level, two authors first conducted open coding on an initial 10\% sample to identify a first approximation of the codebook, along with their prior knowledge of work in the field. In coding papers to these facets, we code the system's operative construct of intent, since the user's mental state cannot be recovered from a paper's description of a system. Both authors then reflected upon this initial codebook to discuss the consistency and comprehensiveness of the categories, where categories were merged, expanded, or removed.


The two authors then independently coded a random sample of 50 papers (25\% of the corpus) on all taxonomy dimensions, discussed the value definitions to resolve systematic differences in interpretation, and recoded independently. Remaining disagreements were resolved by discussion, and the resolved codes are used throughout. Once the codebook and coding scheme were finalized, the remainder of the corpus was single-coded, split between the two authors, with periodic meetings to review ambiguous papers and limit coder drift. 


In the following sections, we present our results and findings of this classification, mark taxonomy dimensions and their values directly within the running text, with exemplar references from the corpus grounding each mechanism we describe. 


%% file: sections/5.analysis.tex
\section{Intent-Centered Taxonomy}

In this section, we introduce the taxonomy based on analysis of the corpus. It comprises three levels below intent itself, illustrated in Figure~\ref{fig:taxonomy}. The six facets and their respective values form the first two levels. The third level comprises the interaction mechanisms through which systems realize each value, which we derived bottom-up from our corpus analysis.

\subsection{Grain}

\dimvalue{Distal} A distal intent constitutes a far end state that the user wants brought about, independent of the means for reaching it. For example, in agentAR, users express ``what" AR application they want, without describing a plan to build it \cite{zhu2025agentar}. This is what systems in our corpus tend to name a goal, an aim, or a need. We identify the following mechanisms through which systems elicit it.

\begin{itemize}
    \item \bemph{Outcome specification.}  The user states characteristics of the desired end state in a single free-form utterance, and how it should be realised is left entirely to the system. In Gensors, a user describes the form of a condition to be noticed, such as a plant needing watering, with no indication of specific evidence that satisfy it \cite{liu2025gensors}. 

    \item \bemph{Goal probing.} The system poses questions and the aim is assembled from the answers, so specification becomes a dialogue rather than a single act of authoring. StepWrite decomposes composition into subtasks and sequentially asks what the text must accomplish before drafting \cite{elalaoui2025stepwrite}, while Adaptive Prompt Elicitation uses visual queries whose answers compile into a prompt \cite{wen2026adaptive}. 

    \item \bemph{Exemplar grounding.} The user deliberately supplies an existing artifact and the system reads the aim off its content rather than off any statement of purpose. NoTeeline treats fragmentary micronotes as evidence of what the user is trying to capture \cite{huq2025noteeline}.

    \item \bemph{Contextual inference.} The system infers the aim from sensed context that the user never submitted as a specification, and often without any expressive act at all as seen in Sensible Agent, which uses egocentric cameras and multimodal sensing to determine both what assistance to offer and how to deliver it \cite{lee2025sensible}.

\end{itemize}

\dimvalue{Proximal} A proximal intent constitutes a concrete operation or a checkable local property of the outcome, closer to what systems in our corpus call a task, an instruction, or a parameter than a goal or an aim. For example, in Scones the user issues a natural-language instruction that directly modifies the current sketch \cite{huang2020scones}. We identify the following mechanisms through which systems in our corpus elicit it.

\begin{itemize}
    \item \bemph{Action specification.} Same as outcome specification but at proximal level as in the case of Scones \cite{huang2020scones} above.

    \item \bemph{Structured specification.} The user expresses a proximal intent by setting values in controls the system provides, either a single widget, a set of fields filled in before execution, or a selection among system-provided options. In EmoShortcuts the only input is an intensity slider that scales an already-recognised emotion \cite{seo2025emoshortcuts}. Because the controls are surfaced by the system, the user can only state steps that the available parameters can represent, and anything outside that set has to be left unsaid. The dependence is clearer where the system builds the controls from the user's own earlier input as in AdaptiveSliders, which generates semantic attribute sliders from a prompt \cite{jain2025adaptivesliders}.

    \item \bemph{Scoped expression.} The user attaches an expression to a bounded region or span of an existing artifact so that it applies only there as in DirectGPT, where users can prompt what must be done while attaching its scope to a certain object by clicking it \cite{masson2024directgpt}.

    \item \bemph{Output editing.} The user modifies the generated artifact itself, and the modification is what expresses the next intent. Textoshop treats generated text as an editable surface where moving, shortening, and expanding fragments stand in for written instructions \cite{masson2025textoshop}. ImaginationVellum lets the user edit the generative strokes the system placed on the canvas, each freehand revision prompting a new version or partial re-draw \cite{marquardt2025imaginationvellum}. 

\end{itemize}

\dimvalue{Motor}
A motor intent constitutes the movement or action that itself carries out the intended act, rather than a sign that stands for an act and must first be decoded.

\begin{itemize}

    \item \bemph{Programming by demonstration.}  Rather than describing what they want done, the user carries out the action itself, and the system takes the performed movements, clicks, and keystrokes as the specification. For instance, SonifyAR captures the AR events a user physically triggers in situ and uses that record as the context from which sound effects are produced \cite{su2024sonifyar}.

\end{itemize}

\subsection{Gap}

\dimvalue{Underconstrained} This gap arises because users left something open, whether deliberately or without considering it. For example, in Sonora a user's voice command to add a sound to the audio world specifies what to add but may leave aspects like volume, spatial placement, and how it blends with the existing soundscape unspecified \cite{delatorre2025sonora}. We identify the following mechanisms through which systems in our corpus close this kind of gap.

\begin{itemize}
    \item \bemph{Default and edit.} The system supplies its own value for whatever the user left open and treats the result as acceptable unless the user changes it as in Sonora, which fills unspecified aspects of a 3D audio world with defaults \cite{delatorre2025sonora}. 

    \item \bemph{Structural decomposition.} Instead of settling the open value, the system exposes the places where it was open. Narrative Motion Blocks derives a slider-based interface from the animation prompt, exposing the quantities the prompt left open as parameters with default and bounded values the animator then sets \cite{bourgault2025narrative}.

    \item \bemph{Probabilistic revision.} The system attaches a confidence estimate to its filling and updates it as further evidence arrives. GUM follows a propose, confidence, revise pipeline over repeated observations of computer use \cite{shaikh2025creating}. This mechanism typically requires the interaction to continue past a single exchange, since revision depends on evidence the system has not yet seen.
\end{itemize}

\dimvalue{Underexpressed} The user has a specific answer in mind and has not stated it, so the system must recover it. For example, in Gensors a user may know exactly what counts as their toddler "getting into mischief" but be unable to state the precise criteria that define it \cite{liu2025gensors}. We identify the following mechanisms through which systems in our corpus do this.

\begin{itemize}
    \item \bemph{Confirm or correct.} The system proposes an interpretation of what the user already has in mind and treats any correction as an error signal about that inference rather than as a new preference. DiLogics requests confirmation when its inferred web action is uncertain, and updates the trace it replays elsewhere on the page once corrected \cite{pu2023dilogics}. Systems present this as an explicit checkpoint, since a wrong inference and a correct one look identical if the user is not asked.
    
    \item \bemph{Closed-Set Selection.} The system reduces the unknown to a small closed set and has the user pick from it, which suits cases where the answer is easier to recognize than to state. VIVID generates four dialogues reflecting the instructor's highlighted scenario, and the instructor compares them and selects the one that best matches the pedagogical intention they already held \cite{choi2024vivid}. While the selection act itself is easy, the set has to contain the answer as the mechanism provides no way for the user to supply one that is missing.

    \item \bemph{Show-to-resolve.} The system accepts an example or an annotated instance in place of a description, for cases where the user can point at the answer but not phrase it. Gensors lets users supply examples or annotated images to close the gap when they know a sensing condition but cannot articulate it \cite{liu2025gensors}. The mechanism implicitly assumes that users have access to an instance during the interaction.

    \item \bemph{Diagnostic probing.} The system asks diagnostic questions aimed at clarifying the gap in the intent, with each question conditioned on what the previous answers ruled out. Automating UI Optimization detects ambiguity in the user's instruction and asks a targeted clarification question, continuing the cycle until every relevant aspect is unambiguous \cite{li2026automating}.

\end{itemize}

\dimvalue{Indeterminate} No answer exists yet, in the system or in the user's own thinking, and the interaction is where one gets built. For example, in ImaginationVellum a designer begins early-stage visual ideation without a fixed concept in mind, arranging sketches and images spatially on a canvas to discover what they are after \cite{marquardt2025imaginationvellum}. This is supported through:

\begin{itemize}
    \item \bemph{Dialogic elicitation.} The system asks open-ended questions and the answers accumulate into a representation of the user's needs. The interaction resolves the user's understanding of what they need. CASEbot guides users through open-ended conversation to turn a vague health question into a structured, well-specified self-experiment \cite{ishita2026casebot}. This mechanism resembles diagnostic probing at the underexpressed gap, but the questions here are exploratory rather than diagnostic.

     \item \bemph{Candidate generation.} The system proposes candidate criteria or outcomes with no or minimal user guidance. PersonaFlow simulates a set of domain-expert personas from a research topic, and the perspectives the user keeps determine whose feedback shapes the idea from then on \cite{liu2025personaflow}. StoryEnsemble generates diverse personas, problems and solutions from minimal input, each selection seeding the next stage \cite{suh2025storyensemble}.

    \item \bemph{Interactive Exploration.} The system generates candidates based on the user's interaction, and what the user picks feeds into the next round, so preference accumulates across rounds. ImaginationVellum ties generation to the sketch path (and proximity to tags) on the canvas, so the path itself produces the output \cite{marquardt2025imaginationvellum}.

\end{itemize}

\subsection{Temporal Scope}

\dimvalue{Momentary} The intent is applied to the current turn and leaves no representation that can reliably influence subsequent interaction as in the case of OmniActions \cite{li2024omniactions}, which infers a digital action from whatever multimodal sensory input the user encounters in the moment, without carrying the inferences forward.

\dimvalue{Decaying} The intent remains available to both users and the system after it is expressed but progressively loses influence as new content, revisions, or evidence accumulate. We identify the following forms in our corpus, distinguished by what causes the decline.

\begin{itemize}
    \item \bemph{Fading intent.} Intent is retained implicitly within the accumulated interaction context, and its effective influence decreases as newer content competes for representation or attention. For example, Scaling Context-Aware Task Assistants tracks a user's procedure from an initial demonstration, and as dialogue and new sensor readings refine step context in real time, the influence of the original demonstration recedes \cite{arakawa2025scaling}. Once older content is displaced, the system can no longer reliably use it to shape outcomes.

    \item \bemph{Embedded intent.} Intent is captured in the artifact itself, so it decays as the artifact is revised. For example, ImaginateAR holds intent solely in the present state of the augmented scene \cite{lee2025imaginatear}, and Texterial treats text itself as a deformable material, so the intent lives in the material's current state and each shaping gesture overwrites the previous one \cite{jshen2026texterial}.

    \item \bemph{Belief intent.} Intent is represented as a maintained belief that is continuously updated as new evidence arrives. Earlier interpretations are revised as their confidence changes rather than displaced from context. GUM repeatedly revises its confidence-weighted account of the user as new activity is observed \cite{shaikh2025creating}.
\end{itemize}

\dimvalue{Persistent} The intent remains available across subsequent interaction and does not inherently lose influence over time. We distinguish:

\begin{itemize}
    \item \bemph{Retrievable intent.} Intent persists as a record the user can return to, but only as a past state to be restored rather than as something that can be modified in place. For example, HistoryPalette keeps every prior version and alternative available to revisit and reuse, as stored past states rather than specifications that can be edited in place \cite{benharrak2026historypalette}.

    \item \bemph{Reified intent.} Intent persists as a representation that users can directly interact with. While some of these representations only allow users to review and revise, such as Stylette turning a styling goal into a palette of CSS properties and values the user can inspect and apply \cite{kim2022stylette}, others such as Malleable User Interfaces \cite{cao2025generative}, go further and turn the representation into an interactive object that can be reused for subsequent interaction.

\end{itemize}

\subsection{Explicitness}

\dimvalue{Explicit} The user states the content of the intent directly, and the system does not need to infer it from context or behaviour. Here, we do not count pronoun disambiguation as implicit. An instruction such as "make it bigger" is explicit about the requested change, even though the system must look at the canvas to find out what "it" refers to. We identify the following mechanisms through which systems in our corpus achieve this.

\begin{itemize}
    \item \bemph{Natural Language.} The full intent, or a self-contained unit of it, is stated in a single written or spoken utterance that the system parses directly. CoPrompter treats the text prompt as encapsulating the entirety of a user's requirements \cite{joshi2025coprompter}, and ViseGPT lets the user specify a data-wrangling script entirely in natural language \cite{zhu2025visegpt}.

    \item \bemph{Direct Manipulation.} Parameters are mapped explicitly to the aspect of the outcome they affect, with no verbal encoding step between the action and the value it sets. Users can directly manipulate these parameters to express their intent. For example, Geppetto lets users edit a robot's motion parameters directly in a semantic space derived from its physics simulation \cite{desai2019geppetto}.

    \item \bemph{Co-expressive.} Language or direct manipulation alone is often insufficient to convey the intent, so systems bind multiple channels together such that the combination carries meaning neither channel carries alone. DirectGPT translates a direct manipulation action, such as dragging, into engineered prompt text, binding the manipulation to the prompt syntax it fires \cite{masson2024directgpt}.

\end{itemize}

\dimvalue{Implicit} Here, the system must infer the content of the intent itself often because it goes unstated and language cannot easily hold it. We identify the following mechanisms through which implicit intent is conveyed to systems.

\begin{itemize}
    \item \bemph{Artifact.} The system infers the aim from an existing artifact the user already has, rather than from anything created to instruct the system. DesignTrace lets the designer drag an image onto a category and extracts the attributes from it, so what the reference is taken to specify is read off the image rather than stated \cite{peng2026designtrace}.

    \item \bemph{Trace.} The system infers the aim from a trace of the user's own activity, whether a continuous path or a discrete mark, made for an immediate purpose other than instructing the system. ImaginationVellum reads sketch path and proximity to tags on a canvas as the basis for generation \cite{marquardt2025imaginationvellum}.

    \item \bemph{Embodied.} This sits at the far end of the implicit intent spectrum. The system infers intent from a low-level record of the user's actions accumulated over time, without the user directing any expression toward the system at all. Key example of this is GUMBO, which infers routines from computer-use logs through a propose-confidence-revise pipeline \cite{shaikh2025creating}.
 
\end{itemize}

\subsection{Provenance}

\dimvalue{User-authored} The user steers the content of the intent, and the system adds no steering of its own regardless of how much processing it does along the way. We identify the following ways in which this holds.

\begin{itemize}
    \item \bemph{Pure transduction.} The system converts what the user already specified into behavior, without querying, or generating anything beyond that conversion. For example, GazePointAR fuses gaze direction, pointing gestures, and speech into a single disambiguated query in one pass \cite{lee2024gazepointar}.

    \item \bemph{Corrective steering.} The system produces content toward an intent the user has already fixed, and the user edits it back into line with what they had in mind. SketchGPT is an example of this, proposing a reading of the user's sketch and instructions followed by a feedback phase that lets the user amend into what they meant \cite{huang2025sketchgpt}. The system's content contribution is large while its intent contribution is nil, which is what keeps provenance user-authored.
    
\end{itemize}

\dimvalue{System-authored} The system steers the content of the intent before the user has taken any turn, so the only steering left to the user is to accept or reject what results. This is typically seen in proactive systems as follows.

\begin{itemize}
    \item \bemph{Inference from observation.} The system infers content from behavior it has observed rather than from anything requested, since acting first means there is no user turn to steer around. PrISM-Observer, for example, watches a user perform a routine procedure through a smartwatch and intervenes when it detects a step has been skipped or done out of order \cite{arakawa2024prism}.

\end{itemize}

\dimvalue{Co-authored} Steering passes between user and system at least once, so the resulting content reflects neither party's initial contribution alone. We identify the following ways control changes hands.

\begin{itemize}
    \item \bemph{Candidate steering.} The system generates a set of options and the user selects among them, so what the user comes to want is constituted by what was offered. A different candidate set would have produced a different intent. CoExploreDS generates candidate problem and solution nodes as the user moves through the design space, and the ones the user accepts determine which direction exploration continues in \cite{chen2025coexploreds}.

    \item \bemph{Proactive steering.} The system actively introduces content to steer the user's intent rather than passively responding to what the user supplies. This can be seen in Navigating the Unknown, where the system elicits constraints and preferences the user "may not have initially known or thought of" \cite{peng2025navigating}.

    \item \bemph{Autonomous completion.} The system settles something the user never constrained, so the user, at times even unknowingly, comes to have their intent steered. agentAR builds and deploys an entire AR application from a spoken description, settling structure and implementation the user never specified \cite{zhu2025agentar}.
    
\end{itemize}

\subsection{Commitment degree} 




The overwhelming majority of our corpus does not differentiate commitment degree at all, and anything the user communicates is acted upon. We therefore describe the values the facet suggests rather than mechanisms observed in the corpus, and return to this gap in Section~\ref{sec:discussion}. A \textbf{Desire} is a pro-attitude the system may weigh, partially satisfy, or set aside, rather than a commitment it must fulfill. An \textbf{Intention}, by contrast, is a commitment the system is expected to satisfy, while the choice of means remains with the system.



%% file: sections/6.discussion.tex
\section{Discussion \& Findings}
\label{sec:discussion}
Applying the taxonomy to our corpus makes visible not only how individual mechanisms address individual facets, but also broader tendencies in how
current systems take up user intent as a whole. We first summarize the overall picture, before discussing three aspects in more detail. 

Most papers address intent across multiple levels rather than as a single representation. Distal and proximal intent form the dominant combination, with outcome specification and structured specification as their respective predominant
mechanisms. Together they form the corpus's central translation pattern. Motor-level intent is rare, and is largely confined to embodied demonstration and correction, suggesting a specialized regime rather than a general third level.

Indeterminate and underconstrained gaps together account for a large part of the corpus. This indicates that the literature is concerned primarily with intent that is not yet sufficiently specified for action, rather than with transmitting an already-formed one. Provenance correspondingly skews co-authored over user-authored or system-authored alone. Candidate steering, proactive steering and autonomous completion are the dominant mechanisms.


Commitment degree is the taxonomy's most underdeveloped facet. In almost all papers, the system does not distinguish a tentative intention from one the user has committed to, or from a concrete plan. This gap is notable given how much of the corpus addresses incomplete or evolving goals, and it exposes a dimension current mechanisms largely leave unspecified.
 
\subsection{Intent Interaction Is Primarily a Translation Problem}

The corpus's strongest pattern is a translation from outcome-level intention to actionable specification. Outcome specification dominates at the distal level and structured specification at the proximal level. This captures a recurring strategy in which the user states what they want and the system converts it into parameters, structure, or executable constraints. Alternative distal mechanisms such as Exemplar grounding are comparatively rare. This points to a consequential assumption, namely that systems are usually built to resolve how an intended outcome should be expressed, rather than whether that outcome has been formed in the first place. This distinction matters most for generative or iterative systems, where the user's initial input may itself be exploratory.

\subsection{The Gap Shapes Both Interaction Strategy and Provenance Mechanisms}

Systems predominantly expand the space of possibilities when intent is indeterminate, structure the specification of underconstrained intent, and validate or repair underexpressed intent. This progression reflects different assumptions about what aspects of intent the user has already settled. Users with no settled target benefit from candidates that help them recognize or reflect, while those with a broad target benefit from decomposition and parametrization. Users with settled targets, on the other hand, primarily benefit from correction or confirmation when they express it (or the system interprets it) imperfectly.

The distribution of intent authorship in our corpus tracks this same structure. In most cases, Candidate steering co-occurs with candidate generation; autonomous completion co-occurs with default-and-edit; corrective steering co-occurs with confirm-or-correct. Provenance, in other words, is aligned with intent resolution: the system contributes candidates to steer unformed intent, completes plausible defaults when a broad target exists with underspecified detail, and takes a corrective role if a target exists but has been poorly expressed.

This reframes the common "user versus system control" framing in human-AI interaction. The relevant question is not how much control each actor has, but what kind of intent the user can supply at that point in the interaction. A system generating candidates for an undecided user and a system autonomously filling in missing parameters may both be called shared-authoring systems, yet they support fundamentally different forms of intent.

\subsection{Interaction Mechanisms Favor Recognition Over Specification}

A broader pattern cuts across these gap-specific strategies. Systems frequently emit an intermediate representation first and ask users to react to it, rather than requiring direct specification of intent. This holds across all three major gap types. System-emitted candidates or representations appear in most underconstrained cases, indeterminate cases, and underexpressed cases. The prevalence of generate-and-react mechanisms reflects not only how intent is distributed but also the affordances of contemporary generative systems, where cheap candidate generation makes showing something the default way to resolve intent rather than asking the user to describe it. This exposes a mismatch between what the user knows and what the interaction asks of them. Recognition is often treated as a universal fallback even when intent is already well formed.

%% file: sections/7.future.tex
\section{Future Opportunities}

Based on our findings, we discuss opportunities to move from designing interactions that merely produce outcomes toward designing interactions that explicitly represent, negotiate, and resolve intent.

\subsection{Making Intent an Interaction Object}

A first opportunity is to make intent itself something users can inspect and manipulate. Our taxonomy shows that persistent intent is already commonly represented in existing systems, allowing users to refine or reuse intent to influence subsequent system behavior. However, most papers use reification as a way to store, access, or refine, rather than as an interactive object that can itself serve as an instrument for acting on other data. Thus, while intent may persist beyond an individual interaction, it is rarely made into a first-class object through which users can directly operate on the system or its artifacts.

The instrumental interaction paradigm offers one way to extend this idea. In this paradigm, objects are not only represented for inspection, but can be manipulated through reusable instruments that mediate actions on other objects. Recent AI systems have begun to explore related forms of interaction through AI instruments ~\cite{riche2025ai} and compositional constructs such as cells, generators, and lenses ~\cite{kim2023cells}. Building on these directions, future systems could treat an intent representation not merely as something to remember or revise, but as an instrument through which users can act on data, artifacts, or other system behavior. Such an intent object could capture not only a desired outcome, but also constraints, preferences, unresolved dimensions, and the evidence underlying an inference. Users could then inspect and modify these aspects and apply the resulting intent across interactions or artifacts, without having to repeatedly communicate the same intent through individual expressions.

This is particularly promising for implicit intent. Our taxonomy suggests that systems can infer intent from behavior or interaction traces, but rarely combine such inference with persistent, inspectable, and user-revisable representations. Making inferred intent visible could create a new interaction loop: the system proposes an interpretation, the user corrects the interpretation itself, and subsequent outputs inherit that correction rather than requiring repeated edits. This shifts interaction from correcting the consequences of an inferred intent to directly shaping the representation that guides subsequent system behavior.

Intent objects could also be represented relationally rather than as isolated states. Representing relations between intents such as \emph{inherits from}, \emph{conflicts with}, or is an \emph{alternative to} could allow systems to reason over the history and structure of intentions, opening opportunities for systems that support not only intent specification but also its ongoing negotiation and evolution.

\subsection{Representing Intent as Uncertain and Evolving}

Making intent an explicit object also raises a fundamental question: an explicit representation need not imply a definitive interpretation. In many interactions, the system does not know exactly what the user intends, and the user may not yet know either. Indeed, the prevalence of indeterminate and underconstrained gaps in our corpus suggests that interaction often begins with incomplete or ambiguous intent. Yet existing mechanisms tend to resolve this uncertainty into a single interpretation that is subsequently acted upon. Conversely, retaining multiple interpretations without distinguishing their relative plausibility could overwhelm users and make intent difficult to manage. The challenge is therefore to represent uncertainty while maintaining a usable account of which intents are currently active, tentative, or no longer relevant.

Maintain intent probabilistically offers a promising direction. Some recent proactive systems have begun to explore this direction: GUM (General User Models) ~\cite{shaikh2025creating}, for example, maintains a persistent, probabilistic model of the inferred intents across interactions. Such models of intent could also allow users to return to earlier interpretations when they become relevant again, shifting intent management from remembering a sequence of user instructions to maintaining an evolving model of the user's goals, including how alternative interpretations have emerged, been revised, and gained or lost relevance over time.

\subsection{Modelling Commitment Degree of Intent}
Uncertainty alone, however, does not capture the user's relationship to an intention. Two users may have equally clear intentions but differ substantially in how committed they are to them. Conversely, a system may be highly confident about an intention that the user is only tentatively exploring. Our taxonomy highlights that current mechanisms rarely distinguish them.

Future interaction mechanisms could therefore model commitment independently from desire to intention, fundamentally changing when an AI should act. For example, a highly likely but weakly committed intention might warrant a preview or suggestion, whereas a strongly committed intention may justify execution. Commitment could also be negotiated through interaction rather than inferred solely from behavior. Users might explicitly mark an intention as provisional, authorize the system to experiment with it, or commit only selected aspects while leaving others open. This provides a basis for more nuanced control over delegation and reversibility. More broadly, modelling commitment would address the gap between understanding intent and acting on intent.

\subsection{Making Provenance Traceable}

Our taxonomy also distinguishes intent by authorship: intent in our corpus could be user-authored, system-authored, or co-authored, each realized through different mechanisms. Co-authored intent, however, is seldom tracked once formed. When a user and a system jointly shape an intention through iterative suggestion and revision, most mechanisms discard the provenance of each contribution, leaving no record of which parts of the resulting intent originated with the user and which with the system. CoexploreDS~\cite{chen2025coexploreds} is among the exceptions, using fill status to record which party authored each part of the intent as it is assembled. 

This lack of tracking carries a further risk. Co-authoring mechanisms that surface system suggestions for user acceptance can, over repeated interaction, produce \emph{induced} intents, where users come to believe that a system's suggestion was their own idea rather than one introduced by the system ~\cite{zhang2024vrcopilot}. Without a record of authorship, users and systems alike lose the ability to distinguish an intention the user brought to the interaction from one the system introduced and the user subsequently adopted, an ambiguity that becomes more consequential as co-authoring mechanisms grow more capable of anticipating and proposing intent on the user's behalf.

Future work could address this gap in two related ways. First, systems could track and convey authorship provenance throughout the co-authoring process, leveraging existing interface properties such as fill status or designing novel ones to indicate, whether each intent originated with the user or the system, and if both, by how much. Second, this provenance record could ground mechanisms for accountability, attributing responsibility for an intent, and for outcomes that follow from acting on it, to the party that introduced it. Together, these mechanisms would keep authorship visible and contestable throughout an interaction, rather than letting it collapse into a single, undifferentiated account of what the user wanted.

\subsection{Adapting Interaction Mechanisms on the Fly}

The final opportunity concerns how systems should adapt the way interact with the user when intent remains uncertain, evolving, or fails to resolve. Our findings show that most systems commit to a single mechanism rather than adapting it when user's needs change. The taxonomy, however, suggests that candidate generation, structured specification, and correction mechanisms address different epistemic states of the user, and future systems could diagnose this state directly and dynamically select the mechanism that matches it. For an uncertain user, the system could enter an exploratory mode and generate diverse candidates; as preferences become clearer, it could transition into a specification mode that identifies unresolved parameters and constraints; and once the user has a stable target but the system's interpretation of it remains uncertain, it could switch to clarification or confirmation rather than generating another complete artifact.

This points towards developing intent-resolution policies analogous to a mixed-initiative dialogue manager, but operating over the structure of intent rather than conversational turns alone. Such a policy could continuously estimate or learn from the user's goal state and ongoing interactions to then determine whether the next interaction should be a question, a candidate, a structured editor, a preview, or an execution. Adapting mechanisms in this way would let systems treat interaction strategy itself as a decision to be revised, rather than a fixed property of the interface chosen once at design time.

%% file: sections/8.conclusion.tex
\section{Conclusion}

We introduced a six-facet taxonomy of user intent in Human-AI co-creation, grounded in philosophical accounts of mind, action, and meaning, and applied it to 198 systems from eleven SIGCHI venues, identifying different kinds of interaction mechanisms. Relating mechanisms to the facets they address reframes the central design question: not just what a system lets a user do, but what facet of intent a mechanism presupposes, elicits, or acts upon. Our findings show that different forms of incomplete intent are associated with distinct interaction and authorship strategies, and mechanisms that look alike at the interface often address fundamentally different problems. The taxonomy gives researchers and designers a shared vocabulary for diagnosing why a mechanism fits, or fails, the intent it must support, and grounds the opportunities we outline for the design of future Human-AI interactive systems.